\documentclass[preprint,12pt]{elsarticle}
\biboptions{comma,sort&compress}

\usepackage{amssymb}
\usepackage{amsthm}

\def\nin{\noindent}
\def\beq{\begin{equation}}
\def\eeq{\end{equation}}
\def\bea{\begin{eqnarray}}
\def\eea{\end{eqnarray}}
\def\nnb{\nonumber}
\def\la{\langle}
\def\ra{\rangle}
\def\ga{\left(}
\def\dr{\right)}

\def\beq{\begin{equation}}
\def\eeq{\end{equation}}
\def\bea{\begin{eqnarray}}
\def\eea{\end{eqnarray}}
\def\bq{\begin{quote}}
\def\eq{\end{quote}}
\def\ve{\vert}
\def\nnb{\nonumber}
\def\ga{\left(}
\def\dr{\right)}

\def\nnb{\nonumber}
\def\la{\langle}
\def\ra{\rangle}
\def\nin{\noindent}
\def\ba{\begin{array}}
\def\ea{\end{array}}

\def\als{\alpha_s}

\def\gg2{ \la\alpha_s G^2 \ra}
\def\gg3{g^3f_{abc}\la G^aG^bG^c \ra}
\def\ggg4{\la\als^2G^4\ra}

\def\beq{\begin{equation}}
\def\enq{\end{equation}}
\def\beqa{\begin{eqnarray}}
\def\enqa{\end{eqnarray}}
\def\nnb{\nonumber}

\def\MeV{\nobreak\,\mbox{MeV}}
\def\GeV{\nobreak\,\mbox{GeV}}

\newcommand{\rag}{\rangle}
\newcommand{\lag}{\langle}

\def\gg{\lag g^{2}_{s} G^2 \rag}
\def\ggg{\lag g^{3}_{s}G^3\rag}

\journal{Journal of Subatomic Particles and Cosmology}

\begin{document}

\begin{frontmatter}



\title{$2^{++}$ Di-gluonium from LSR at higher order\tnoteref{invit}}
\tnotetext[invit]{Talk given at the 14th International Conference in High-Energy Physics  - HEPMAD24 (21-26th October 2024, Antananarivo-Madagascar).}

\author[label1]{Siyuan Li}
\ead{siyuan.li@usask.ca}
\address[label1]{Department of Physics \& Engineering Physics, University of Saskatchewan, SK, S7N 5E2, Canada}
\author[label2]{Stephan Narison}
\ead{snarison@yahoo.fr}
\address[label2]{Laboratoire
Univers et Particules de Montpellier (LUPM), CNRS-IN2P3, Case 070, Place Eug\`ene
Bataillon, 34095 - Montpellier, France}
\author[label3]{Davidson Rabetiarivony\fnref{fn2}}
\fntext[fn2]{Speaker}
\ead{rd.bidds@gmail.com}
\address[label3]{Institute of High-Energy Physics of Madagascar (iHEPMAD), University of Antananarivo, Antananarivo 101, Madagascar}
\author[label1]{Tom Steele}
\ead{tom.steele@usask.ca}
\begin{abstract}
\noindent
We improve the determination of the mass and coupling of the $2^{++}$ tensor di-gluonium by using relativistic QCD Laplace sum rules (LSR). In so doing, we evaluate the next-to-leading order (NLO) corrections to the perturbative (PT) and $\lag \alpha_s G^2 \rag$ condensate and the lowest order (LO) $\lag G^3 \rag$ contributions to the $2^{++}$ di-gluonium two-point correlator. Within a vacuum saturation estimate ($k_G=1$) of the dimension-eight gluon condensates, we obtain: $M_{T}=3028(287)\MeV$ and the renormalization group invariant (RGI) coupling $\hat{f}_T=224(33)\MeV$. Assuming that the factorization hypothesis can be violated, we study the effect of the violation factor $k_G$ on the results and  obtain: $M_T=3188(337)\MeV$ and $\hat{f}_T=245(32)\MeV$ for $k_G=(3\pm 2)$. Our estimation does not favour the interpretation of the observed $f_2(2010)$, $f_2(2300)$ and $f_2(2340)$ as pure glueball state.
\end{abstract}
\end{frontmatter}
\section{Introduction}
In earlier papers\,\cite{ANRls,ANR21,ANR1,SU3}, we have used the inverse Laplace Transform of QCD spectral sum rules (QSSR)\,\cite{SNB1,SNB2} to estimate the masses and decay constants of some light, heavy-light and doubly heavy molecule and tetraquark states.

\nin In this talk, based on the paper in Ref\,\cite{LNSR}, we improve the LO estimated mass and coupling of the $2^{++}$ tensor di-gluonium\,\cite{NSVZ,SNG0,SNB2,SNG} by including NLO corrections to the PT and $\lag \alpha_s G^2\rag$ contributions to the correlation function and studying the effect of the violation of factorization hypothesis for dimension $D=8$ gluon condensates\,\cite{SVZ} on the results.
\section{The two-point correlation function}
We shall work with the two-point function:
\bea
\psi^{\mu\nu\rho\sigma}_{T}(q^2)&\equiv & i \int d^4 x\, e^{i q x}\lag 0 \ve \mathcal{T} \theta^{\mu\nu}_G(x)(\theta^{\rho\sigma}_G(0))^{\dag} \ve 0 \rag \nnb \\
&=&\bigg(P^{\mu\nu\rho\sigma} \equiv \eta^{\mu\rho}\eta^{\nu\sigma}+\eta^{\mu\sigma}\eta^{\nu\rho}-\frac{2}{n-1}\eta^{\mu\nu}\eta^{\rho\sigma} \bigg)\psi_{T}(q^2),
\eea
built from the gluon component of the energy momentum tensor:
\beq
\theta^{\mu\nu}_{G}=\alpha_s\left[-G^{\mu,a}_{\alpha}G^{\nu\alpha}_{a}+\frac{1}{4}G^{a}_{\alpha\beta}G^{\alpha\beta}_{a}\right],
\eeq
with:
\beq
\eta^{\mu\nu}\equiv g^{\mu\nu}-q^{\mu}q^{\nu}/q^2 ~~ \mbox{and} ~~ P_{\mu\nu\rho\sigma}P^{\mu\nu\rho\sigma}=2(n^2-n-2),
\label{eq:projct}
\eeq
where: $n=4+2\epsilon$ is the space time dimension used for dimensional regularization and renormalization.
\section{QCD expression of the two-point function}
\subsection{Lowest Order contribution}
The PT and gluon condensates up to dimension $D=8$ contributions to the correlation function reads:
\bea
\psi_{T}\vert_{LO}(Q^2)\hspace*{-0.1cm}=\hspace*{-0.1cm} a^{2}_{s}\left[-\frac{Q^4}{20}\log \ga \frac{Q^2}{\nu^2}\dr +\frac{\pi^2}{6\alpha_s}\lag \alpha_s G^2\rag + \frac{5}{3Q^4} \pi^3 \alpha_s \lag 2O_1-O_2 \rag \right],
\label{eq:lo}
\eea
with: $a_s=\alpha_s/\pi$ and $O_1=(f_{abc}G_{\mu\alpha}G_{\nu\alpha})^2\; , O_2=(f_{abc}G_{\mu\nu}G_{\alpha\beta})^2.$
Assuming vacuum saturation hypothesis ($k_G=1$), the $D=8$ gluon condensates can be expressed as:
\beq
\lag 2O_1-O_2\rag \simeq -k_G\left(\frac{3}{16}\right)\lag G^2 \rag^2 ,
\eeq
where $k_G$ is the violation factor.\\
Compared to the expression obtained in Ref.\,\cite{NSVZ}, we found a non-zero $\lag \alpha_s G^2 \rag $ term. However, this term does not affect the analysis as it will disappear when one takes the derivatives of the two-point function.
\subsection{Next-to-Leading Order contribution}
\subsubsection*{$\bullet$ PT expression}
We shall be concerned with the bare diagrams listed in Table\,1 of Ref.\,\cite{LNSR}. We perform the calculation in two ways: diagrammatic renormalization and conventional operator renormalization methods. The diagrammatic renormalization for QCD sum rules correlation functions has been initiated and developed in Ref.\,\cite{STEELE}. The conventional renormalization approach uses the standard Feynman approach\,\cite{SNB1,SNB2} by considering the renormalization of the gluonic current $\theta^{\mu\nu}_{G}/\alpha_s$ with the renormalization constant in Ref.\,\cite{BAGAN}. The diagrammatic and conventional renormalization approaches lead to the same result. The $\overline{MS}$ renormalized two-point correlation function corrected to NLO for $n_f$ flavours reads:
\beq
\psi^{pert}_{T}\vert^{R}_{NLO}(Q^2)=\psi^{pert}_{T}\vert_{LO}\bigg[1+a_s\bigg(\frac{n_f}{6}\log\ga \frac{Q^2}{\nu^2}\dr -\frac{101 n_f+150}{90}\bigg)\bigg],
\label{eq:pt-nlo}
\eeq
where the PT LO term $\ga\psi^{pert}_{T}\vert_{LO}\dr$ can be deduced from Eq.\,\ref{eq:lo}. In the case of gluodynamics, we reproduce the results of Ref.\,\cite{PIVO}.
\subsubsection*{$\bullet$ Dimension-four gluon condensate}
We use the renormalization group equation (RGE) to evaluate the leading-log term of $\lag \alpha_s G^2 \rag$ at NLO\,\cite{SNB4,ASNER}. One obtains\,\cite{LNSR}:
\beq
\psi_{T}^{G^2}\vert_{NLO}(Q^2)=\frac{1}{24\pi} \alpha^{2}_{s} \lag \alpha_s G^2 \rag \ga 11+\frac{2 n_f}{3}\dr \log\ga\frac{Q^2}{\nu^2}\dr .
\label{eq:GG-nlo}
\eeq
\section{$2^{++}$ Di-gluonium mass and coupling at NLO}
\subsection*{$\bullet$ The Laplace sum rules}
To estimate the mass and decay constant, we shall use the finite energy version of Laplace sum rule and their ratios:
\beq
{\cal L}^{c}_{0,1}(\tau,\nu)\equiv  \int_{t_>}^{t_c} dt~t^{(0,1)}~e^{-t\tau}\frac{1}{\pi} \mbox{Im}~\tilde{\psi}_{T}(t,\nu)~;~~ {\cal R}^c_{10}(\tau)\equiv \frac{{\cal L}^c_{1}} {{\cal L}^c_0}.
\label{eq:lsr}
\eeq
The contribution of the di-gluonium state to the spectral function can be introduced within the Minimal Duality Ansatz (MDA):
\beq
\frac{1}{\pi} \mbox{Im}~\tilde{\psi}_{T}(t)=f^{2}_{T}M^{4}_{T}\delta(t-M^{2}_{T})+\theta(t-t_c)"QCD~continuum",
\eeq
where $f_T$ is normalized as $f_{\pi}=132\MeV$. In the MDA:
\beq
{\cal R}^c_{10}\simeq M^{2}_{T}.
\eeq
Combining Eqs.\,\ref{eq:lo}, \ref{eq:pt-nlo} and \ref{eq:GG-nlo}, one can deduce the LO$\oplus$NLO QCD expression of the correlator up to $D=8$ gluon condensates and normalized to $\alpha^{2}_{s}$: $\tilde{\psi}_T(Q^2)$.
The moment ${\cal L}^{c}_{0}$ is evaluated by taking the third (3rd) derivative of the correlator $\tilde{\psi}_T(Q^2)$ while the ${\cal L}^{c}_{1}$ moment is obtained by taking the forth (4th) derivative of $Q^2\tilde{\psi}_{T}(Q^2)$. The expressions of the moments ${\cal L}^{c}_{0,1}$ for $n_f=3$ flavours to order $\alpha_s$ and up to dimension-8 condensates are:
\bea
{\cal L}_{0}^{c}&=&\frac{\tau^{-3}}{10\pi^2}\left\lbrace \Big{[} 1-a_s\ga \frac{53}{15}+\gamma_E\dr\Big{]}\,\rho^{c}_{2} -\frac{65\pi}{12}\la\alpha_s G^2\ra\tau^2\rho_{0}^{c} \right. \nnb \\
&&\left. - \frac{\pi^2}{a_s}\ga \frac{25}{8}\dr k_G\la\alpha_s G^2\ra^2\tau^4 \right\rbrace,
\eea
and
\bea
{\cal L}_{1}^{c}&=&\frac{3\tau^{-4}}{10\pi^2}\Bigg{\{}\Bigg{[} 1-a_s\ga \frac{16}{5}+\gamma_E\dr\Bigg{]}\,\rho^{c}_{3}-\frac{65\pi}{36}\la\alpha_s G^2\ra\tau^2\rho_{1}^{c}\nnb\\
&& + \frac{\pi^2}{a_s}\ga \frac{25}{24}\dr k_G\la\alpha_s G^2\ra^2\tau^4\Bigg{\}},
\eea
with $\gamma_{E}=0.5772\cdots$ is the Euler constant and 
\beq
\rho_{n}^{c}=1-e^{-t_c\tau}\ga 1+(t_c\tau)+\cdots+\frac{(t_c\tau)^n}{n!} \dr .
\eeq
\subsection*{$\bullet$ Optimization criteria}
We shall use stability criteria on the external variables $\tau$ and $t_c$ for extracting the optimal results from the sum rules. The stability regions manifest either as minimum or inflexion point in $\tau$ while we shall take $t_c$ in a conservative region from the beginning of $\tau-$stability until the $t_c-$stability. The subtraction constant $\nu$ is eliminated when one takes the different derivatives of the correlator to obtain the inverse Laplace transform and working with the running QCD parameters.
\subsection*{$\bullet$ Mass at NLO}
\subsubsection*{-- Vacuum saturation hypothesis}
\begin{figure}[hbt]
\begin{center}
\centerline { \bf \hspace*{-6.5cm} a) \hspace*{6.0cm} b)}
\includegraphics[width=6.8cm]{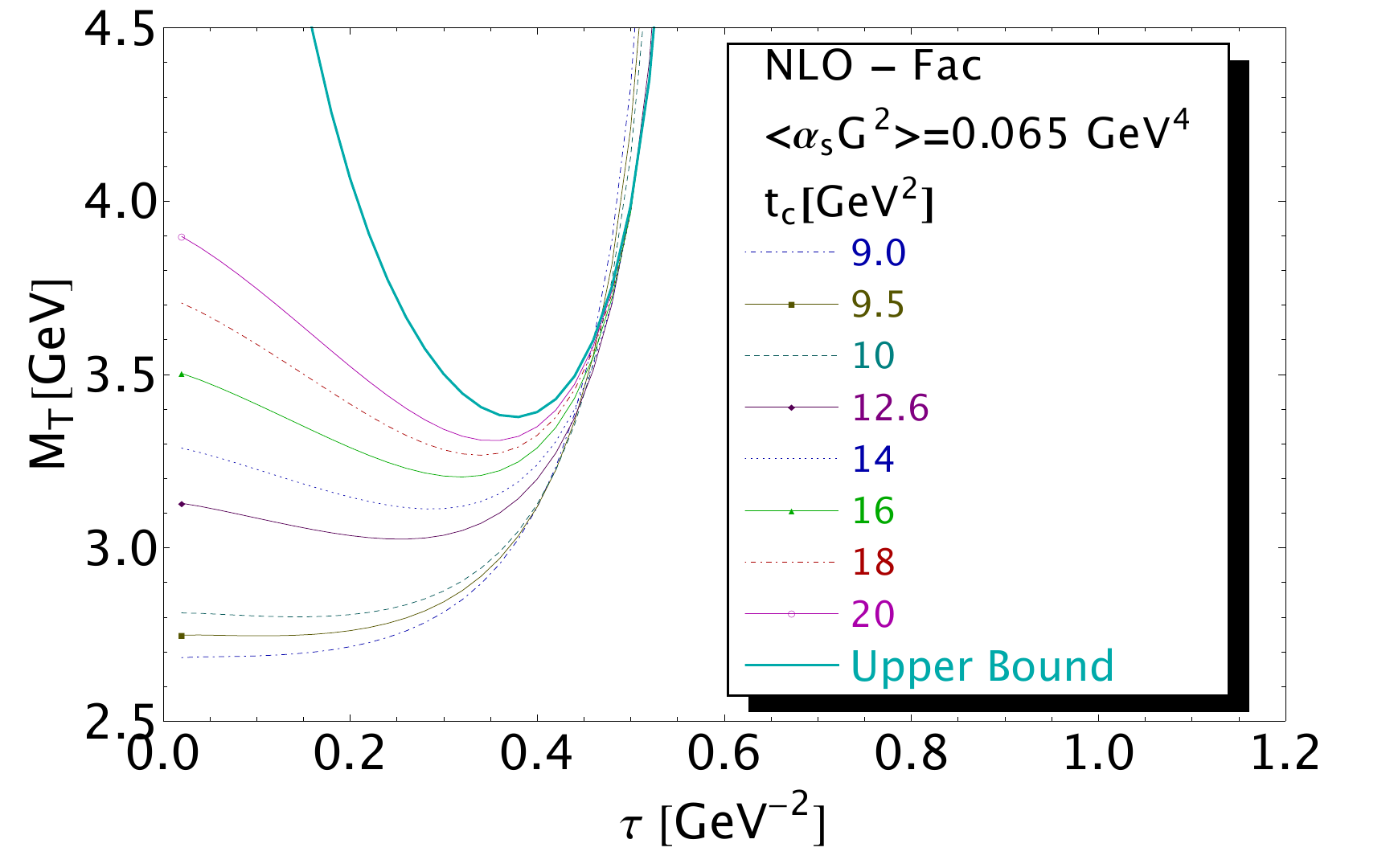}
\includegraphics[width=6.7cm]{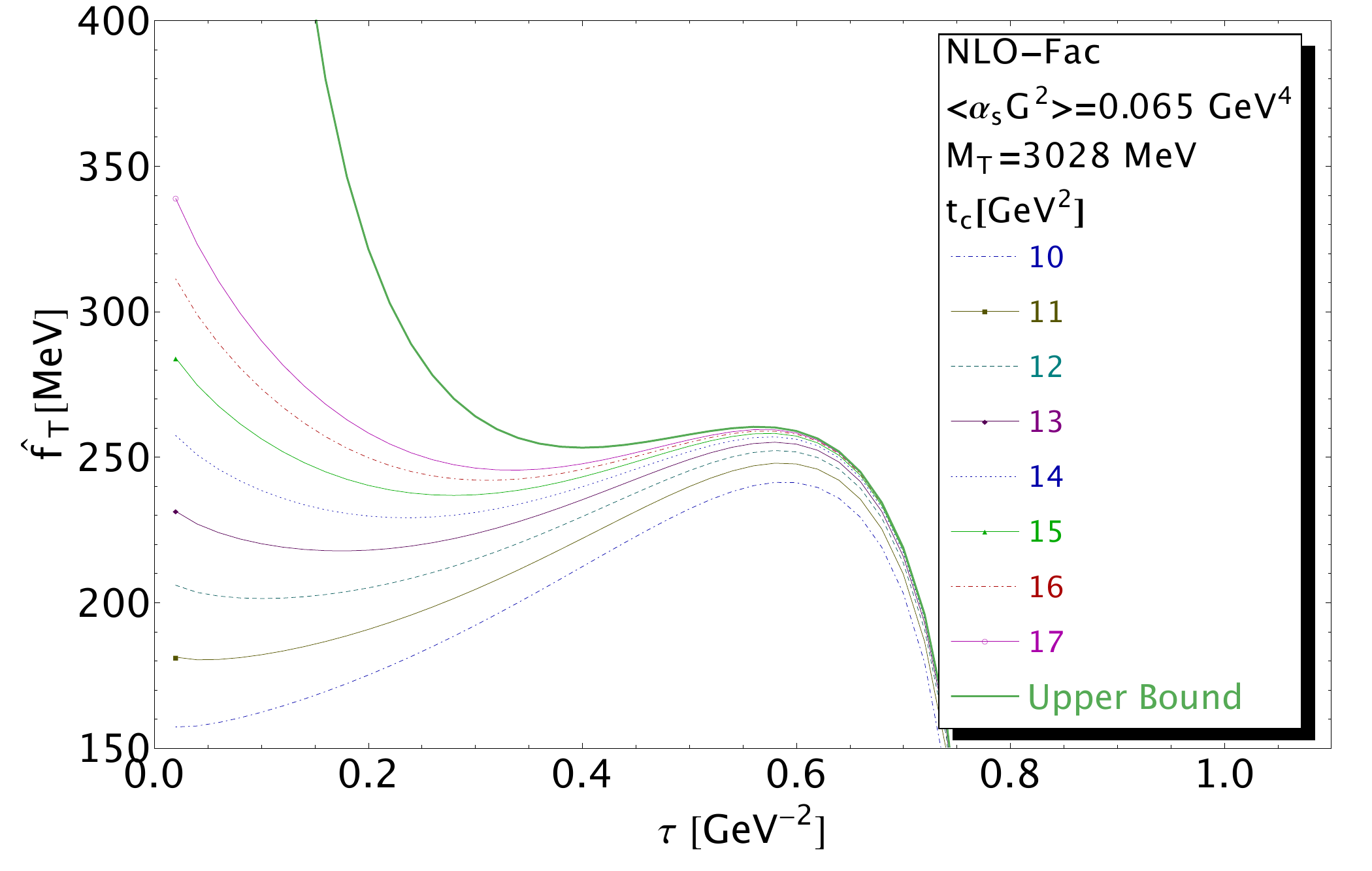}
\caption{\footnotesize a) $\tau-$behaviour of the $2^{++}$ di-gluonium mass at NLO for different values of $t_c$ where the factorization of $D=8$ gluon condensates is assumed. b) The same caption as a) but for the coupling $\hat{f}_T$.} 
\label{fig:fac-nlo}
\end{center}
\end{figure} 
The behaviour of the mass at NLO and assuming factorization ($k_G=1$) of the dimension-8 gluon condensates is shown in Fig.\,\ref{fig:fac-nlo}-{\bf a)}. The optimal values of the mass are extracted for the set $(\tau,t_c)=(0.12,9.5)\sim(0.36,20)\,(\GeV^{-2},\GeV^2)$ where one obtains respectively: $2746$ and $3309\MeV$. These values lead to the mean:
\beq
\lag M_T\rag = 3028(287)\MeV \rightsquigarrow t_c \simeq 12.6 \GeV^2 .
\eeq
The optimal upper bound from the positivity of the ratio of moments is:
\beq
M_T\leqslant \left[ 3376(26)_{\Lambda}(42)_{G^2} = 3376(49)\right]\MeV.
\eeq
\subsubsection*{-- Violation of vacuum saturation hypothesis}
Assuming that the vacuum saturation hypothesis can be violated ($k_G\neq 1$), we study the effect of the dimension $D=8$ gluon condensates on the mass determination. Varying the value of the violation factor $k_G$ from $1$ to $5$, the mass increases slightly of about $+319\MeV$. Like in the case of the four-quark condensate, we shall use the conservative range $k_G=(3\pm 2)$ to extract the optimal value of the mass. We obtain:
\beq
\lag M_T\rag = 3188(337)\MeV .
\eeq
\subsection*{$\bullet$ Coupling at NLO}
Noting that, the current $\theta^{\mu\nu}_{G}$ acquires an anomalous dimension due to its renormalization\,\cite{LNSR}; thus we have to introduce the renormalization group invariant (RGI) coupling $\hat{f}_{T}$ which is related to the running coupling $f_{T}(\nu)$ as:
\beq
f_{T}(\nu)=\frac{\hat{f}_{T}}{\ga \log\ga \frac{\nu}{\Lambda}\dr \dr^\frac{\gamma_1}{-2\beta_1}}~;~~\gamma_1=\frac{n_f}{3}.
\eeq
We shall extract the coupling from the first moment ${\cal L}^{c}_{0}$.
\subsubsection*{-- Factorization of $D=8$ gluon condensates ($k_G=1$)}
At NLO, as shown in Fig.\,\ref{fig:fac-nlo}-{\bf b)}, the minimum and inflexion point of the curves are more pronounced. The optimal values are extracted at the $\tau-$minimum which are respectively $201$ and $246\MeV$ for the set $(\tau,t_c)=(0.1,12)$ to $(0.34,17)\,(\GeV^{-2},\GeV^2)$. We obtain the mean:
\beq
\hat{f}_T\ve_{NLO} = 224(33)\MeV \rightsquigarrow t_c \simeq 13.5 \GeV^2.
\eeq
The optimal upper bound is extracted at the minimum $\tau=0.4\GeV^{-2}$:
\beq
\hat{f}_T\ve_{NLO} \leqslant 253(1)_{\Lambda}(2)_{G^2}(32)_{M_T}=253(32)\MeV.
\eeq
\subsubsection*{-- Violation of factorization of $D=8$ gluon condensates}
We use the value of $k_G=(3\pm2)$ to perform the analysis. The behaviour of the curves are similar to the ones in Fig.\,\ref{fig:fac-nlo}-{\bf b)}. The optimal values $225$ (resp. $265$)\,$\MeV$ are obtained for the set of $(\tau,t_c)$: $(0.14,14)$ (resp. $(0.36,22)$) $(\GeV^{-2},\GeV^2)$. They lead to the mean:
\beq
\hat{f}_T\ve_{NLO} = 245(32)\MeV \rightsquigarrow t_c \simeq 7.5 \GeV^2.
\eeq
At the inflexion point $\tau=0.36\GeV^{-2}$, we extract the optimal upper bound:
\beq
\hspace*{-0.4cm} \hat{f}_T\ve_{NLO} \leqslant 268(3)_{\Lambda}(1.5)_{G^2}(30)_{M_T}=268(32)\MeV.
\eeq
\section{Comments and conclusion}
We have presented improved predictions of QSSR of the mass and coupling of $2^{++}$ tensor di-gluonium state using LSR method. We have evaluated the PT and $D=4$ gluon condensate NLO corrections and tested the effect of the estimate of the $D=8$ gluon condensates on the results.\\
-- The PT (resp. $\lag \alpha_s G^2 \rag$) NLO corrections increase the mass by about $561$ (resp. $376$) $\MeV$ from its LO value: $2091\MeV$\,\cite{LNSR}.\\
-- Moving the mass from its LO value to the NLO one: $3028\,\MeV$, the $\alpha_s$ corrections increase the value of the coupling by about $84\MeV$. However, if one fixes the mass at its LO value, the coupling is slightly affected by the NLO corrections.\\
-- Our results for NLO $\lag \alpha_s G^2 \rag$ and LO $\lag G^3 \rag$ condensates term do not agree with the ones in Ref.\,\cite{ZHUG}. The discrepancy may come from the different current used by the authors in\,\cite{ZHUG}.\\
-- Compared to other approaches, the predicted mass of lattice QCD which is expected to be in the range $(2.27\sim2.67)\GeV$\,\cite{LATTG06,LATTG12,TEPER} is slightly lower than ours and the one of instanton liquid model ($1525\MeV$) is too low. However, our estimated mass goes in line within the errors with the ones of ADS/QCD\,\cite{ADS} and some constituent models\,\cite{VENTO}.\\
-- Confronted to experimental data, our prediction does not favour the interpretation of $f_2(2010)$, $f_2(2300)$ and $f_2(2340)$ states as pure glueball or meson-gluonium mixing\,\cite{BAGAN} candidates.

\end{document}